\documentclass[letterpaper, 10 pt, conference]{ieeeconf}
\IEEEoverridecommandlockouts  
\usepackage{microtype}
\usepackage{graphicx}
\usepackage{subfigure}
\usepackage{booktabs} 

\usepackage{hyperref}

\usepackage[utf8]{inputenc} 
\usepackage[T1]{fontenc}    

\usepackage{url}            
\usepackage{booktabs}       
\usepackage{amsfonts}       
\usepackage{nicefrac}       
\usepackage{microtype}      
\usepackage{amsmath}       
\usepackage{mathtools}
\usepackage{blindtext}
\usepackage{centernot}
\usepackage{algorithm}
\usepackage{algorithmic}
\usepackage[algo2e]{algorithm2e} 
\usepackage{subfigure}
\usepackage{svg}
\usepackage{bbm}

\DeclareMathOperator*{\argmax}{arg\,max}

\DeclarePairedDelimiterX{\infdivx}[2]{(}{)}{%
  #1\;\delimsize\|\;#2%
}

\newtheorem{corollary}{Corollary}
\newtheorem{theorem}{Theorem}

\newtheorem{definition}{Definition}
\newtheorem{assumption}{Assumption}

\newenvironment{thmproof}[1]{\par\noindent\textit{Proof of #1: }}{\hfill$\square$\par}

\def\underbracex#1#2{\mathop{\vtop{\m@th\ialign{##\crcr
   $\hfil\displaystyle{#2}\hfil$\crcr
   \noalign{\kern3\p@\nointerlineskip}%
   #1\crcr\noalign{\kern3\p@}}}}\limits}

\def\upbracefilla{$\m@th \setbox\z@\hbox{$\braceld$}%
  \bracelu\leaders\vrule \@height\ht\z@ \@depth\z@\hfill 
\kern\p@\vrule \@width\p@\kern\p@\vrule \@width\p@\kern\p@\vrule \@width\p@
$}

\def\upbracefillbd{$\m@th \setbox\z@\hbox{$\braceld$}%
\vrule \@width\p@\kern\p@\vrule \@width\p@\kern\p@\vrule \@width\p@\kern\p@
\bracerd\braceld
 \leaders\vrule \@height\ht\z@ \@depth\z@\hfill\braceru$}

\allowdisplaybreaks

\usepackage[textsize=tiny]{todonotes}
\setuptodonotes{inline}

\title{\LARGE \bf
Approximately Solving Continuous-Time Mean Field Games with Finite State Spaces 
}

\author{
Yannick Eich$^{1}$,
Christian Fabian$^{1}$,
Kai Cui$^{2}$,
Heinz Koeppl$^{1}$%
\thanks{$^{1}$Department of Electrical Engineering and Information Technology, 
Technische Universität Darmstadt, Germany (e-mail: yannick.eich@tu-darmstadt.de).}
\thanks{$^{2}$Huawei Technologies Co., Ltd., China.}}
\begin{document}

\maketitle
\thispagestyle{empty}
\pagestyle{empty}

\begin{abstract}
Mean field games (MFGs) offer a powerful framework for modeling large-scale multi-agent systems. This paper addresses MFGs formulated in continuous time with discrete state spaces, where agents' dynamics are governed by continuous-time Markov chains - relevant to applications like population dynamics and queueing networks. While prior research has largely focused on theoretical aspects of continuous-time discrete-state MFGs, efficient computational methods for determining equilibria remain underdeveloped. 
Inspired by discrete-time approaches, we approximate the classical Nash equilibria by regularization methods, enabling more computationally tractable solution algorithms. Specifically, we define regularized equilibria for continuous-time MFGs and extend the classical fixed-point iteration and fictitious play algorithm to these equilibria.
We validate the effectiveness and practicality of our approach via illustrative numerical examples.

\end{abstract}

\section{Introduction}
Learning equilibria in multi-agent systems is crucial for many applications in control theory but remains computationally challenging as the number of agents increases \cite{daskalakis2009complexity, deng2023complexity}. Mean field games (MFGs) \cite{huang2006large,lasry2007mean} offer a scalable framework, maintaining fixed complexity regardless of the number of agents, by exploiting the exchangeability and symmetry among agents.

While MFGs have been extensively studied, most existing literature focuses either on discrete-time formulations or on continuous-time formulations with continuous state spaces. In contrast, this work adresses continuous-time finite-state MFGs, where each agent's dynamics follow a continuous-time Markov chain (CTMC). Such CTMC-based formulations are practically significant, as many real-world processes, including biological systems~\cite{wilkinson2018stochastic}, queueing networks~\cite{bolch2006queueing}, and epidemiological models~\cite{britton2010}, naturally evolve in continuous time over discrete state spaces. 

Modeling these processes directly in continuous time has several well-established practical and numerical advantages. Although time discretization is a common approach, it is known to introduce sensitivity issues and inaccuracies in inherently continuous-time dynamics, as extensively discussed in~\cite{tallec2019making, jia2023q}.
By contrast, continuous-time approaches naturally align with the underlying dynamics and support adaptive step sizes, enabling computationally efficient and numerically robust algorithms.
 
As a result, recent literature has begun exploring mean field control and games under CTMC dynamics.
Cooperative mean field control has been considered in various works \cite{cecchin2021finite, bauerle2023continuous, wei2023continuous}. Closer to our setting, previous work on competitive continuous-time finite-state MFGs has provided their theoretical foundation \cite{gomes2013continuous, bayraktar2018analysis, cecchin2019convergence}, under fixed target distributions \cite{averboukh2023planning}, or with conditionally independent agents \cite{bayraktar2022finite}. 

These recent works have primarily focused on the Nash equilibrium (NE) as the central solution concept of the MFG.
However, the computation of NE remains difficult in the general case, motivating the exploration of alternative equilibrium notions as solution concepts. In particular, regularized equilibria (RE) have been introduced in discrete-time MFGs as a tractable alternative to NE, and are known to approximate NE while offering convergence properties and computational advantages under suitable regularization parameters~\cite{cui2021approximately,eich2025bounded}. 
Building on these developments, we extend the notion of regularized equilibria to continuous-time finite-state MFGs.
Specifically, our contributions are as follows:
\begin{itemize}
    \item We introduce the concept of RE for MFGs with CTMC dynamics;
    \item We derive a fixed-point characterization of RE via the entropy-regularized Hamilton-Jacobi-Bellman equation;
    \item We extend the known fixed-point iteration and fictitious play algorithms to compute RE in continuous-time;
    \item We evaluate the effectiveness of our algorithms through comprehensive numerical experiments.
\end{itemize}

\section{MFGs in Continuous Time with finite state spaces} 
In this section, we first introduce finite-agent games evolving in continuous time and their associated MFG formulations. Subsequently, we formally define two solution concepts: the classical NE and the newly proposed RE. These equilibria serve as the desired results for the multi-agent equilibrium learning algorithms that we subsequently introduce. We use the framework of relaxed control (e.g., \cite{piunovskiy2020continuous}), which generalizes deterministic actions to probability measures over the control space. This approach can be implemented using discrete sampled controls \cite{kim2022maximum}, effectively allowing agents to randomize their actions.

\textit{Notation: $\mathcal P(\mathcal X)$ denotes the space of probability measures over a finite set $\mathcal X$, equipped with the $L_1$ norm $\lVert \cdot \rVert$ unless noted otherwise. Products of metric spaces are equipped with the $\sup$ metric. For $N \in \mathbb N$, define $[N] \coloneqq \{ 1, \ldots, N \}$ .}

\subsection{Finite Game}
We consider the finite $N$-agent game of practical interest, where each agent $i \in [N]$ is characterized by random state $x^i_t \in 
\mathcal X$ and action $u^i_t \in \mathcal U$, evolving continuously  over time $t \in [0, T]$, $T \in \mathcal T \coloneqq \mathbb R_{\geq 0}$. The sets $\mathcal X$ and $\mathcal U$ denote the finite state and action spaces, respectively, which are identical across all agents. We denote the empirical mean field by $\mu_t^N \coloneqq \frac{1}{N} \sum_{i=1}^N \mathbf 1_{x_t^i}$, which represents the empirical distribution of the agents' states at time $t$. 
The dynamics of each agent $i$ is described by a controlled CTMC, whose evolution for a small time step $h$ is characterized by
\begin{multline*}
    \mathbb{P}(x^i_{t+h}=x'\mid x^i_t=x, u^i_t=u, \mu^N_t = \nu) \\
    = \delta_{xx'} + h \Lambda(x,x',u,\nu) + o(h),
\end{multline*}
where $\Lambda(x,x',u,\nu)$ denotes the jump rate from state $x$ to $x'\neq x$, depending on the agents action $u$ and the current mean field $\nu$, and where  $\Lambda(x,x,u,\nu) = -\sum_{x' \neq x} \Lambda(x,x',u,\nu)$. Here, $o(h)$ represents terms for which $\lim_{h \to 0} o(h) / h =0$.

Using the relaxed control framework and Markov policies, i.e. letting actions depend on states via chosen policies $\pi^i$ from the set of (measurable in time) policies $\Pi \subseteq \mathcal P(\mathcal U)^{\mathcal X \times [0, T]}$, we obtain the relaxed control dynamics
\begin{multline*}
    \mathbb{P}(x^i_{t+h}=x'\mid x^i_t=x, \mu^N_t = \nu) \\
    = \delta_{xx'} + h \sum_{u \in \mathcal U} \Lambda(x,x',u,\nu) \pi^i_t(u \mid x) + o(h).
\end{multline*}

Each competitive agent $i$ aims to optimize their individual objective while anticipating the behavior of other agents. The exact form of this objective depends on the chosen equilibrium notion (see Section~\ref{sec:notions}), but is typically defined through a reward function that is maximized  cumulatively over time.

\subsection{Mean Field Game}
MFGs arise as the limiting case of finite $N$-agent games as $N \to \infty$, offering a tractable approximation for many-agent finite games. By the law of large numbers, the stochastic empirical mean field $\mu^N_t$ converges to a deterministic limiting mean field $\mu \coloneqq (\mu_t)_{t \in [0, T]} \in \mathcal M \subseteq \mathcal P(\mathcal X)^{[0, T]}$, where $\mathcal M$ is the space of all obtainable mean fields. 
This convergence property enables MFGs to simplify the challenging task of finding approximate (symmetric) equilibria, which is typically difficult in large finite-agent games~\cite{deng2023complexity}.
In this limiting regime, the dynamics of a representative agent $i$ with policy $\pi$, interacting with the mean field $\mu$, are given by
\begin{multline*}
    \mathbb{P}(x^i_{t+h}=x'\mid x^i_t=x, \mu_t = \nu) \\
    = \delta_{xx'} + h \sum_{u \in \mathcal U} \Lambda(x,x',u,\nu) \pi_t(u \mid x) + o(h).
\end{multline*}

 At equilibrium, all agents are assumed to symmetrically play the same policy $\pi \in \Pi$, which constitutes the equilibrium solution.  
The resulting deterministic mean field then evolves according to the master equation
\begin{align} \label{eq:master}
     \frac{\mathrm d\mu_t(x)}{\mathrm dt} = \sum_{u \in \mathcal U} \sum_{x'} \Lambda(x',x,u,\mu_t) \mu_t(x') \pi_t(u \mid x'),
\end{align}
with fixed initial $\mu_0$. Given a policy $\pi$, we write $\mu=\Gamma_{\mathcal M}(\pi)$ for the mean field obtained as the solution of the master equation~\eqref{eq:master}.

\subsection{Equilibria in continuous-time finite-state MFGs} \label{sec:notions}
With MFGs formally defined, we now introduce the relevant notions of equilibria. Several equilibrium concepts exist, each capturing distinct agent behaviors and varying significantly in terms of analytical and computational tractability. We begin by briefly revisiting the classical NE, widely studied in literature. Subsequently, as a primary contribution of this paper, we introduce and analyze RE for CTMC MFG as a computationally attractive alternative to NE.
\paragraph{Nash equilibria}
The common mean field Nash equilibrium (NE) has been considered before in continuous-time finite state MFGs \cite{gomes2013continuous, cecchin2019convergence}. The objective function for a deviating agent with policy $\hat\pi$ is given by
\begin{align}\label{eq:Nash_obj}
    J(\hat \pi, \pi) = \mathbb{E} \bigg[ \int_0^T \sum_{u \in \mathcal U} r(x_s,u,\mu_s) \hat \pi_s(u \mid x_s) \mathrm ds + q(x_T) \bigg]
\end{align}
for running rewards $r \colon \mathcal X \times \mathcal U \times \mathcal P(\mathcal X) \to \mathbb R$ and terminal reward $q \colon \mathcal X \to \mathbb R$, where all other agents choose policy $\pi$ and the mean field follows from $\mu=\Gamma_{\mathcal M}(\pi)$.

Assuming rational agents that predict other agents' decisions, all agents should adopt policies from which no agent can deviate to improve their objective.
The corresponding limiting NE is therefore a policy that is optimal when played against itself, i.e. when all agents employ the same policy.

\begin{definition}[Mean field NE]
    A mean field NE is a policy $\pi^* \in \Pi$ such that $\pi^* \in \argmax_{\pi \in \Pi} J(\pi, \pi^*)$.
\end{definition}
Mean field NE are known to provide approximate NE in large finite $N$-agent games, in the sense that the exploitability is negligible, rigorously motivating MFGs and mean field NE under mild continuity assumptions of the game \cite{gomes2013continuous}.
To characterize and compute NE, we consider the associated stochastic control problem. For a given $\pi$ and resulting mean field $\mu$, the solution to optimizing $J$ is well known \cite[Thm.~3.4]{prieto2012selected}. Applying the principle of optimality, we define the value function $V^{\mu}$ as the solution to the  Hamilton-Jacobi-Bellman (HJB) equation:
\begin{align} \label{eq:ctmc-V} 
    -\frac{\partial}{\partial t} V^\mu_t(x) = \sup_{u \in \mathcal U}  Q_t^{\mu}(x,u)
\end{align}
with terminal condition $V^\mu_T(x) = q(x)$ and with the action value function
\begin{align} \label{eq:ctmc-Q}
 Q_t^{\mu}(x,u)= r(x, u, \mu_t) %
      + \sum_{x'} \Lambda(x,x',u,\mu_t) V^{\mu}_t(x').   
\end{align}
The optimal policy $\hat \pi$ is then obtained from the action value function as:
\begin{align} \label{eq:ctmc-pi-opt}
    \hat \pi_t(x) \in \argmax_{u \in \mathcal U} Q_t^{\mu}(x,u).
\end{align}
Note that while we optimize over deterministic actions in the HJB equation above, the resulting policy is a deterministic (pure) strategy. Within the more general relaxed control framework introduced earlier, this corresponds to choosing a degenerate distribution, concentrated at the optimizing action.

\paragraph{Regularized equilibria}
Computing NE - that is, finding fixed points of the master equation~\eqref{eq:master} along with the optimal policy derived from Eqs.~\eqref{eq:ctmc-V} and \eqref{eq:ctmc-pi-opt} is hard in general. Motivated by discrete-time approaches \cite{cui2021approximately}, we introduce a class of regularized equilibria (RE) to improve computational tractability. 
In this framework, we modify the standard objective in Eq.~\eqref{eq:Nash_obj} by incorporating an entropy regularization term. Specifically, agents are not only incentivized to maximize their expected reward but also to favor more stochastic policies by maximizing the entropy of their control distribution. The entropy of a policy $\hat \pi_t(\cdot \mid x_t)$ is defined as
\begin{align*}
    \mathcal H\left(\hat \pi_t(\cdot \mid x_t)\right) = - \sum_{u \in \mathcal U} \hat \pi_t(u \mid x_t) \log \hat \pi_t(u \mid x_t).
\end{align*}
With a regularization temperature parameter $\alpha>0$, the resulting objective becomes
\begin{equation}
\begin{aligned}
\label{eq:CTRE}
    J^{\mathrm{RE}}_\alpha(\hat \pi, \pi) = \mathbb{E} \left[ \int_0^T  \sum_{u \in \mathcal U} r(x_s,u,\mu_s) \hat \pi_s(u \mid x_s) 
    \right.\\\left.
    + \alpha \mathcal H(\hat \pi_s(\cdot \mid x_s))  \mathrm ds + q(x_T) \vphantom{\sum_{u\in \mathcal U}}\right].
\end{aligned} 
\end{equation} 
Similar to the classical NE setting, we can derive the corresponding entropy-regularized optimality equations with stochastic control theory.
We begin by defining the soft value function $V^{\alpha,\mu}$ as
\begin{equation*}
\begin{aligned}
     V^{\alpha,\mu}_t(x) = \sup_{\hat{\pi} _{[t,T]}} \mathbb{E} \left[ \int_t^T  \sum_{u \in \mathcal U} r(x_s,u,\mu_s) \hat \pi_s(u \mid x_s)\right. \\
     +\left. \alpha \mathcal H(\hat \pi_s(\cdot \mid x_s)) \mathrm ds + q(x_T) \vphantom{\sum_{u\in \mathcal U}}\right].
\end{aligned}
\end{equation*}
Applying the principle of optimality, this optimization can be decomposed over a small interval $[t,t+h]$
\begin{equation}
\begin{aligned}
\label{eq:value_definition_dp}
     V^{
     \alpha,\mu}_t(x) = \sup_{\hat{\pi}_{[t,t+h]}} \mathbb{E} \left[ \int_t^{t+h}  \sum_{u \in \mathcal U} r(x_s,u,\mu_s) \hat \pi_s(u \mid x_s) \right.
     \\
    \left.+ \alpha \mathcal H(\hat \pi_s(\cdot \mid x_s))\mathrm ds +  V^{\alpha,\mu}_{t+h}(x_{t+h})\vphantom{\sum_{u\in \mathcal U}} \right].
\end{aligned}
\end{equation}

Considering the dynamics of the CTMC, the expected soft value function at time $t+h$ can be expanded using the generater of the process as
\begin{equation}
\label{eq:next_value}
\begin{aligned}
    &\mathbb{E} \left[ V^{\alpha,\mu}_{t+h}(x_{t+h}) \right] =  V^{\alpha,\mu}_t(x) + h \frac{\partial}{\partial t}  V^{\alpha,\mu}_t(x) \\
    &+ h \sum_{x'} \sum_{u \in \mathcal U} \Lambda(x,x',u,\mu_t)\hat{\pi}_t(u \mid x) V^{\alpha,\mu}_t(x')  + o(h).
\end{aligned} 
\end{equation}


Substituting Eq.~\eqref{eq:next_value} into Eq.~\eqref{eq:value_definition_dp}, we obtain the optimality condition
\begin{equation*}
\begin{aligned}
   0 =&  \sup_{\hat{\pi}_{[t,t+h]}} \mathbb{E} \left[ \int_t^{t+h}  \sum_{u \in \mathcal U} r(x_s,u,\mu_s) \hat \pi_s(u \mid x_s)\right.\\
   &\left.+ \alpha \mathcal H(\hat \pi_s(\cdot \mid x_s))\mathrm ds \vphantom{\sum}\right]+ h \frac{\partial}{\partial t}  V^{\alpha,\mu}_t(x)\\
   & + h \sum_{x'} \sum_{u \in \mathcal U} \Lambda(x,x',u,\mu_t)\hat\pi_t(u \mid x_t)V^{\alpha,\mu}_t(x') + o(h).
\end{aligned}
\end{equation*}

Dividing by $h$ and taking the limit as $h \to 0$, we arrive at the entropy-regularized HJB equation
\begin{align*}
- \frac{\partial}{\partial t} V^{\alpha,\mu}_t(x)
= \sup_{\hat{\pi}_t} \Biggl[&
    \sum_{u \in \mathcal U}  Q^{\alpha,\mu}_t(x,u)  \hat \pi_t(u \mid x) \\
&+ \alpha \mathcal H\bigl(\hat \pi_t(\cdot \mid x)\bigr)
\Biggr],
\end{align*}
where the soft action value function $Q^{\alpha,\mu}$ is defined analogously to Eq.~\eqref{eq:ctmc-Q}, but now based on the soft value function $V^{\alpha,\mu}$ instead of $V^{\mu}$.

It is known \cite{kim2022maximum} that the optimal policy $\hat \pi^*$ solving this entropy-regularized problem is given by the softmax distribution
\begin{align}\label{eq:ctmc-pi-soft}
    \hat\pi^\ast_t(u\mid x)= \frac{\exp \left[ \frac{1}{\alpha}  Q^{\alpha,\mu}_t(x,u)  \right]}{\sum_{u' \in \mathcal U} \exp \left[ \frac{1}{\alpha}  Q^{\alpha,\mu}_t(x,u')  \right]}.
\end{align}
We denote the mapping from soft action-value function $Q^{\alpha,\mu}$ to the associated softmax policy by $\pi = \Gamma_{\Pi}(Q^{\alpha,\mu})$.
Moreover, the supremum in the HJB equation can be expressed explicitly as
\begin{align*}
\begin{aligned}
    &\sup_{\hat{\pi}_t} \left[ \sum_{u \in \mathcal U}   Q^{\alpha,\mu}_t(x,u) \hat \pi_t(u \mid x)  + \alpha \mathcal H(\hat \pi_t(\cdot \mid x)) \right]\\
    &\qquad =\alpha \log \sum_{u \in \mathcal U} \exp \left[ \frac{1 }{\alpha} Q^{\alpha,\mu}_t(x,u)\right].
 \end{aligned}
\end{align*}

Thus, solving the optimality equations
\begin{align}\label{eq:ctmc-V-soft}
\begin{aligned}
    &0 = \frac{\partial}{\partial t} V^{\alpha,\mu}_t(x) + \alpha \log \sum_{u \in \mathcal U} \exp \left[ \frac{1}{\alpha} Q^{\alpha,\mu}_t(x,u) \right]\\
    &V^{\alpha,\mu}_T(x) = q(x)
\end{aligned}  
\end{align}
we obtain the optimal policy via $\pi = \Gamma_{\Pi}(Q^{\alpha,\mu})$. Given a mean field $\mu$, let $Q^{\alpha,\mu} = \Gamma_{Q}(\mu)$ denote the soft action value function by first solving the HJB Eq.~\eqref{eq:ctmc-V-soft} and then computing $Q^{\alpha,\mu}$ as in Eq.~\eqref{eq:ctmc-Q}.
We can now define RE in MFGs. 

\begin{definition}[RE] \label{def:cteq}
    A RE in continuous-time finite-state MFGs is a policy $\pi^* \in \Pi$ with induced mean field $\mu^*$ such that, $\pi^* = \Gamma_{\Pi}(\Gamma_Q(\Gamma_{\mathcal M}(\pi^*))).$
\end{definition}
The equilibria policies are thus a fixed point of the master equation~\eqref{eq:master} and the HJB equation~\eqref{eq:ctmc-V-soft}.

We show that such equilibria are well-defined, as they are guaranteed to exist under standard Lipschitz conditions.

\begin{assumption} \label{ass:ccont}
    The rates $\Lambda \colon \mathcal X \times \mathcal X \times \mathcal U \times \mathcal P(\mathcal X) \to \mathbb R$ and rewards $r \colon \mathcal X \times \mathcal U \times \mathcal P(\mathcal X) \to \mathbb R$ are Lipschitz continuous.
\end{assumption}

\begin{theorem} \label{thm:ctmc-exist}
    For any $\alpha > 0$, a RE exists under Assm.~\ref{ass:ccont}.
\end{theorem}
As the temperature parameter $\alpha$ approaches zero, RE recover NE, as the entropy-regularized objective converges to the classical NE objective given by Eq.~\eqref{eq:Nash_obj}. Therefore, RE can be viewed as principled approximations of NE.

\section{Learning Non-Cooperative Mean Field Equilibria}\label{sec:algorithms}
Having formally introduced the notion of RE in continuous-time finite-state MFGs, we now present computational methods designed for finding such equilibria. 
Due to the implicit fixed-point characterization of RE, it is natural to employ iterative fixed-point algorithms. In particular we generalize the classical fixed-point iteration (FPI) and fictitious play (FP) algorithm to the RE in our continuous-time setting.
\begin{algorithm}[b]
    \caption{Continuous-time Fixed-Point Iteration for RE.}
    \label{alg1}
    \begin{algorithmic}[1]
        \STATE Input: Temperature $\alpha >0$, initial policy $\pi^0$.      
        \FOR {$k=0,1,\ldots,K-1$}
            \STATE Compute mean field  
            $\mu^k\leftarrow\Gamma_{\mathcal{M}}(\pi^k)$ induced by $\pi^k$.
            \STATE Solve the HJB equation $Q^{k}\leftarrow \Gamma_Q(\mu^k)$.
           \STATE Compute the new softmax policy $\pi^{k+1} \leftarrow 
           \Gamma_{\Pi}(Q^k)$.
        \ENDFOR
    \STATE \textbf{return} 
        $\pi^K$ .
    \end{algorithmic}
\end{algorithm}

\paragraph{Fixed Point Iteration.}
The FPI method proceeds by repeatedly applying the fixed-point operator in Definition~\ref{def:cteq} to update the policy. Starting from an initial policy (e.g. $\pi^{0}_t(u \mid x) = 1 / |\mathcal U|$) the algorithm iteratively computes the resulting mean field, solves the entropy-regularized HJB equation, and finally computes the updated policy via the softmax operator, as
\begin{align*}
    \pi^{k+1} = \Gamma_\Pi(\Gamma_{Q}(\Gamma_{\mathcal M}(\pi^k))).
\end{align*}
 The continuous-time FPI algorithm, presented in Alg.~\ref{alg1}, converges under Lipschitz continuity assumptions for sufficiently high temperatures. 

\begin{theorem} \label{thm:fpi-lip}
In continuous-time finite-state MFGs under Assm.~\ref{ass:ccont}, the fixed-point maps in Def.~\ref{def:cteq} are $L_\alpha$-Lipschitz for some $L_\alpha > 0$ with $L_\alpha \to 0$ as $\alpha \to \infty$.
\end{theorem}

\begin{corollary}
    By Banach's fixed-point theorem, the regularized FPI algorithm converges to a unique equilibrium for sufficiently large $\alpha > 0$.
\end{corollary}

Although strong regularization guarantees convergence, increasing $\alpha$ moves the obtained equilibrium away from the original NE, resulting in equilibria that poorly approximate NE. Therefore, a general learning algorithm is desired.

\begin{algorithm}[b!]
    \caption{Continuous-time Fictitious Play for RE.}
    \label{alg2}
    \begin{algorithmic}[1]
        \STATE Input: Temperature $\alpha > 0$, initial policy $\pi^0$, $\beta \in (0, 1)$.
        \STATE Initialize mean field $\mu^0\leftarrow\Gamma_{\mathcal M}(\pi^0)$ induced by $\pi^0$.
        \FOR {$k=0,1,\ldots,K-1$}
        \STATE Solve the HJB equation $Q^k \leftarrow \Gamma_{Q}(\mu^k)$.
        
        \STATE Compute the softmax policy $\pi^{k+1}\leftarrow\Gamma_{\Pi}(Q^k)$.
  
        \STATE Compute mean field $\mu^{k+1}\leftarrow\Gamma_{\mathcal{M}}(\pi^{k+1})$.

        \STATE Average mean field $\mu^{k+1} \leftarrow (1-\beta)\mu^{k+1} + \beta \mu^k $.
        \ENDFOR

        \STATE \textbf{return}  $\pi^K$.
    \end{algorithmic}
\end{algorithm}

\paragraph{Fictitious Play.}
As an alternative to the strongly regularized FPI method, we propose a generalization of the FP algorithm \cite{perrin2020fictitious}.
Unlike FPI, which directly iterates the fixed-point operator, FP maintains historical averages of previous mean fields. This historical averaging allows FP to achieve convergence at lower temperatures compared to FPI, thus providing equilibria closer to the original NE. The continuous-time FP algorithm is given in Alg.~\ref{alg2}.

\section{Experiments} \label{sec:exp}
In the following, we evaluate our algorithms on three different MFGs: (i) a continuous-time version of the Left-Right (LR) problem \cite{cui2021approximately}, (ii) randomly generated MFGs with a swarm avoiding reward function, and (iii) an epidemic control setting modeled by a Susceptible-Infectious-Susceptible (SIS) problem. Detailed descriptions of these systems are provided in the appendix. All ODEs are solved numerically using a fourth-order Runge–Kutta method with step size $\Delta t = 0.01$.

To assess the performance of our algorithms, we quantify the distance to a RE, similar to the exploitability in the NE case.

\begin{definition}[Distance to equilibria]
    The distance of a policy $\pi \in \Pi$ to a NE and RE are defined as
    \begin{align*}
     \Delta J(\pi)&\coloneqq \max_{\hat \pi \in \Pi} J(\hat \pi, \pi) - J(\pi, \pi),\\
     \Delta J^{RE}(\pi)&\coloneqq \max_{\hat \pi \in \Pi} J^{RE}(\hat \pi, \pi) - J^{RE}(\pi, \pi).\\
    \end{align*}
\end{definition}
A policy $\pi$ is a NE or RE if and only if $\Delta J(\pi) = 0$ or $\Delta J^{RE}(\pi) = 0$, respectively.

\begin{figure}[t]
    \centering
    \includegraphics{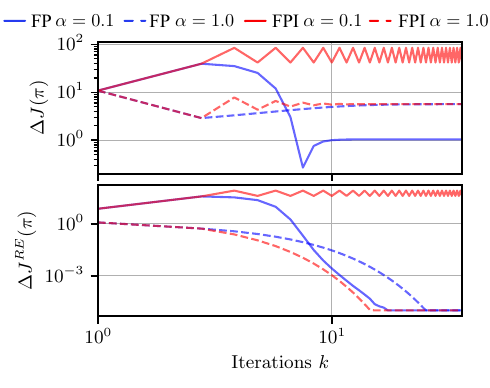}
    \caption{Convergence of FP and FPI on the LR problem.}
    \label{fig:LR}
\end{figure}

\begin{figure}[t]
    \centering
    \includegraphics{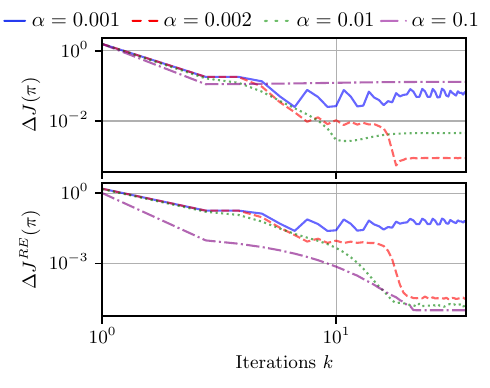}
    \caption{Convergence of FP on a randomly generated MFG.}
    \label{fig:random}
\end{figure}

\begin{figure}[t]
    \centering
    \includegraphics{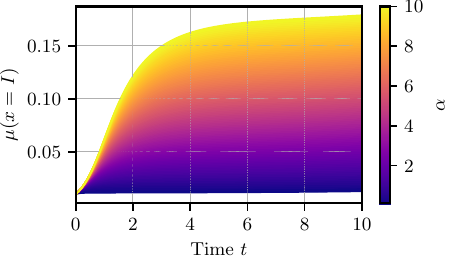}
    \caption{Fraction of infectious agents in SIS RE over temperatures $\alpha$.}
    \label{fig:SIS}
\end{figure}

First, we evaluate the FP and FPI algorithms on the LR problem for different temperature parameters $\alpha=0.1$ and $\alpha=1.0$. Figure~\ref{fig:LR} illustrates the convergence behavior in terms of $\Delta J$ and $\Delta J^{RE}$ over iterations $k$. At $\alpha = 1.0$, both FP and FPI converge successfully to the RE, demonstrating contraction properties of the fixed-point map at higher temperatures. In contrast, for $\alpha = 0.1$, FPI fails to converge, exhibiting oscillatory behavior, whereas FP continues to converge. This highlights the robustness of FP at lower temperatures. Further, we see that the lower $\alpha$ parameter leads to better NE approximations, since the resulting RE has lower exploitability.

Next, we evaluate FP on a randomly generated MFG under varying temperature parameters $\alpha$ to examine the algorithm's ability to approximate NE via RE solutions. As shown in Figure~\ref{fig:random}, FP converges for $\alpha = 0.1$, $0.01$, and $0.002$, with lower temperatures leading to reduced exploitability. For $\alpha = 0.001$, however, FP fails to converge, suggesting that the convergence becomes more difficult as the RE approaches the NE. This highlights a trade-off: while lower $\alpha$ values bring the RE closer to the NE, they also reduce algorithmic stability. Nevertheless, FP provides high-quality approximations even at relatively low temperatures.

Finally, we investigate the SIS epidemic model to study how RE policies affect global agent behavior. For a range of temperatures $\alpha \in [0.1, 10.0]$, we solve the RE using FP and measure the resulting mean field, particularly the average fraction of infected agents in equilibrium.
Figure~\ref{fig:SIS} shows that at lower temperatures, where the equilibrium policy is close to the Nash policy, the infection rates stay small over the time horizon. Conversely, at higher temperatures, which correspond to more heavily regularized and thus less optimal policies, a larger portion of the population gets infected. This demonstrates how proximity to NE can significantly influence emergent system behavior in practical applications like epidemic control.

\section{Conclusion and Discussion}
We introduced RE for continuous-time finite-state MFGs, providing a tractable alternative to NE in settings where direct computation of NE is challenging. By formulating RE through the solution of an entropy-regularized HJB equation, we derived a fixed-point characterization suitable for algorithmic learning.
Building upon this foundation, we generalized the classical FPI and FP algorithms to compute RE effectively in the continuous-time setting and analyzed their convergence behavior. 
Our numerical results highlight the trade-off between approximation quality and algorithmic stability, and show that FP, in particular, maintains robustness across a wide temperature range. 
Future research directions include formal convergence proofs for FP and the development of adaptive regularization techniques.

\section*{Acknowledgements}
This work has been co-funded by the LOEWE emergenCITY research promotion program of the federal state of Hessen, Germany, by the German Research Foundation (DFG) within the Collaborative Research Center (CRC) 1053 MAKI and project number 517777863, by the Federal Ministry of Education and Research as part of the Software Campus project RL4MFRP (funding code 01IS23067) and by the Hessian Ministry of Science and the Arts (HMWK) within the projects "The Third Wave of Artificial Intelligence - 3AI" and hessian.AI.

\appendix
\subsection{Proofs}
\begin{thmproof}{Theorem~\ref{thm:ctmc-exist}}
    We show that for any $\alpha > 0$, the range of $\Gamma^\alpha_\Pi \circ \Gamma_{Q} \circ \Gamma_{\mathcal M}$ is a subset of all $L_\Pi$-Lipschitz policies for some $L_\Pi > 0$.
    To see this, we first note that any $\mu$ generated by any policy is Lipschitz in time, due to the master equation \eqref{eq:master} and its bound $2\Lambda_{\mathrm{max}}$, where $\Lambda_{\mathrm{max}} = \sup \Lambda < \infty$ by Assm.~\ref{ass:ccont}. Accordingly, the same is true for the value functions $Q=\Gamma_Q(\mu)$. Now recall that for any $\mu$, the resulting policy is $\pi = \Gamma^\alpha_\Pi(Q)$. Further, the soft value function and its time derivative by \eqref{eq:ctmc-V-soft} are bounded for any $(t, \mu)$. The spaces $[0, T]$ and $\mathcal M$ (of $2\Lambda_{\mathrm{max}}$-uniformly Lipschitz mean fields) are compact \cite[Lemma~2]{cui2023learning}. There hence exists a global bound $L_{V}$ for \eqref{eq:ctmc-V-soft}, which is a global Lipschitz constant of $V$. Lastly, by the definition of policies \eqref{eq:ctmc-pi-soft}, we immediately obtain a global Lipschitz constant of action values $Q$ and policies $\pi$ from $L_{V}$. 
    Finally, the space of $L_\Pi$-Lipschitz policies is convex and closed under the supremum norm, as limits of sequences of policies are still measurable and $L_\Pi$-Lipschitz \cite[Lemma~2]{cui2023learning}. Therefore, the existence follows by Schauder's fixed point theorem \cite[Thm.~5.28]{rudin1991functional}.
\end{thmproof}

\begin{thmproof}{Theorem~\ref{thm:fpi-lip}}
    Recall that we equip $\mathcal P(\mathcal X)^{[0, T]}$ with the uniform norm, i.e. for $\mu, \mu' \in \mathcal P(\mathcal X)^{[0, T]}$ let
    \begin{align*}
        d(\mu, \mu') = \sup_{t \in [0, T]} \lVert \mu_t - \mu'_t \rVert.
    \end{align*}
    Analogously, the space of policies $\Pi \subseteq [0, 1]^{[0, T] \times \mathcal X \times \mathcal U}$ is equipped with the uniform norm, i.e. for any $\pi, \pi' \in \Pi$ we write $\pi(t, x, u) = \pi_t(u \mid x)$ and have
    \begin{align*}
        d(\pi, \pi') = \sup_{(t, x, u) \in [0, T] \times \mathcal X \times \mathcal U} | \pi(t, x, u) - \pi'(t, x, u) |.
    \end{align*}

    We write $L_\Lambda$, $L_r$ for the Lipschitz constants of $\Lambda$ and $r$ respectively by Assm.~\ref{ass:ccont}.

    \paragraph{Lipschitz $\Gamma_{\mathcal M}$.} 
    First, we show that the map $\Gamma_{\mathcal M}$ is Lipschitz under appropriate conditions. Consider any $\pi, \pi'$, then let $\mu = \Gamma_{\mathcal M}(\pi), \mu' = \Gamma_{\mathcal M}(\pi')$ and note that we let
    \begin{align*}
        d(\mu, \mu') = \sup_{t \in [0, T]} \lVert \mu_t - \mu'_t \rVert.
    \end{align*}

    Now as in \cite{branicky1994continuity}, define $F \colon [0, T] \times \mathbb R^{\mathcal X} \to \mathbb R^{\mathcal X}$ such that for all $t \in \mathcal T, \nu \in \mathcal P(\mathcal X)$, the $x$-th component is
        \begin{align*}
        F(t, \nu)_{x} \coloneqq \sum_{u \in \mathcal U} \sum_{x'} \Lambda(x',x,u,\nu) \nu(x') \pi(t,x',u),
    \end{align*}
    and analogously define $G \colon [0, T] \times \mathbb R^{\mathcal X} \to \mathbb R^{\mathcal X}$ with $\pi'$.
    Observe that, for $F$ and $G$, we have for all $t,\nu,\nu'$
    \begin{align*}
        \lVert F(t, \nu) - F(t, \nu') \rVert &\leq 2 |\mathcal X| (|\mathcal X| - 1) (L_\Lambda + \Lambda_{\mathrm{max}}) \lVert \nu - \nu' \rVert
    \end{align*}
    and similarly for all $t,\nu$
    \begin{align*}
        \lVert F(t, \nu) - G(t, \nu) \rVert \leq |\mathcal X| (|\mathcal X| - 1) \Lambda_{\mathrm{max}} d(\pi, \pi').
    \end{align*}
    
    Therefore, by $\dot \mu_t = F(t, \mu_t)$, $\dot \mu'_t = G(t, \mu'_t)$ and \cite[Lemma~2 with $A=0, c=1, \eta=0$] {branicky1994continuity},
    \begin{align*}
        &\lVert \mu_t - \mu_t' \rVert \leq \sqrt{|\mathcal X|} \lVert \mu_t - \mu_t' \rVert_2 \\
        &\leq \sqrt{|\mathcal X|} \frac{|\mathcal X| (|\mathcal X| - 1) \Lambda_{\mathrm{max}} d(\pi, \pi')}{2 |\mathcal X| (|\mathcal X| - 1) (L_\Lambda + \Lambda_{\mathrm{max}})} \\
        &\cdot\left( \exp \left( (2 |\mathcal X| (|\mathcal X| - 1) (L_\Lambda + \Lambda_{\mathrm{max}})) t \right) - 1 \right)
    \end{align*}
    at all times $t$, and hence we have the Lipschitz constant $L_{\Gamma_{\mathcal M}}$ as
    \begin{align*}
        &d(\mu, \mu')
        \leq
       \frac{\Lambda_{\mathrm{max}} \sqrt{|\mathcal X|}}{2 (L_\Lambda + \Lambda_{\mathrm{max}})}\\
        &\cdot
        \left( \exp \left( (2 |\mathcal X| (|\mathcal X| - 1) (L_\Lambda + \Lambda_{\mathrm{max}})) T \right) - 1 \right) d(\pi, \pi').
    \end{align*}

    \paragraph{Lipschitz $\Gamma^\alpha_{\Pi} \circ \Gamma_{ Q}$.} 
    Next, we show that the map $\Gamma^\alpha_{\Pi} \circ \Gamma_{ Q}$ is Lipschitz. Consider any $\mu, \mu'$, then for $\pi = \Gamma^\alpha_{\Pi}(\Gamma_{Q}(\mu)), \pi' = \Gamma^\alpha_{\Pi}(\Gamma_{Q}(\mu'))$ we have
        \begin{align*}
       & d(\pi, \pi')
        = \sup_{t \in [0, T]} \sup_{(x, u) \in \mathcal X \times \mathcal U} | \pi(t, x, u) - \pi'(t, x, u) | \\
        &= \left| \frac{\exp \left[ \frac{1}{\alpha}   Q^{\alpha,\mu}_t(x,u)   \right]}{\sum_{u'} \exp \left[ \frac{1}{\alpha}  Q^{\alpha,\mu}_t(x,u')  \right]} 
        - \frac{\exp \left[ \frac{1}{\alpha}   Q_t^{\alpha,\mu'}(x,u)  \right]}{\sum_{u'} \exp \left[ \frac{1}{\alpha}  Q_t^{\alpha,\mu'}(x,u') \right]} \right|.
    \end{align*}
    Note that if we can show that $V^{\alpha,\mu}_t(x)$ is Lipschitz in $\mu$ with uniform constant $L_V$, then the map 
        \begin{align*}
        \mu \mapsto r(x,u,\mu_t) + \sum_{x'} \Lambda(x,x',u,\mu_t)  V^{\alpha,\mu}_t(x') 
    \end{align*}
    is Lipschitz with constant $L_r + 2 |\mathcal X| \Lambda_{\mathrm{max}} L_V$. By \cite[Lemma~B.7.5]{cui2021approximately}, we then have the desired conclusion
    \begin{align*}
        d(\pi, \pi') &\leq \frac{|\mathcal U| - 1}{2\alpha} (L_r + 2 |\mathcal X| \Lambda_{\mathrm{max}} L_V).
    \end{align*}

    It hence only remains to show that $ V^{\alpha,\mu}_t(x)$ is Lipschitz in $\mu$ with uniform constant $L_V$. Note that by inverting the optimality ODE \eqref{eq:ctmc-V-soft} in time, we can use the same arguments as for ${\Gamma_{\mathcal M}}$. More precisely, define $W^{\alpha,\mu}_t \coloneqq  V^{\alpha,\mu}_{T-t}$ at all times $t$, and note that the Lipschitz constants of $W$ and $V$ in $\mu$ are equal. Then, $W$ follows the reversed ODE
    \begin{align*}
        \partial_t W^{\alpha,\mu}_t(x) = \alpha \log \sum_{u \in \mathcal U} \exp \left[ \frac{r(x,u,\mu_{T-t}) + \Lambda[W_t^{\alpha,\mu}]}{\alpha} \right],
    \end{align*}
    with $\Lambda[W_t^{\alpha,\mu}]=\sum_{x'} \Lambda(x,x',u,\mu_{T-t}) W^{\alpha,\mu}_t(x')$.
    Defining again $F^\alpha$ and $G^\alpha$ as the right-hand sides, given $\mu$ and $\mu'$ respectively, we then have
    \begin{align*}
        \dot W^{\alpha,\mu}_t = F^\alpha(t, W^{\alpha,\mu}_t), \quad \dot W^{\alpha,\mu'}_t = G^\alpha(t, W^{\alpha,\mu'}_t),
    \end{align*}
    \vspace{0.5cm}
    with
      \begin{align*}
        &\lVert F^\alpha(t, w) - F^\alpha(t, w') \rVert \\
        &\quad \leq |\mathcal X| \sup_x \left| \frac{\sum_{u \in \mathcal U} \exp \left[ \xi_u \right] \cdot \Lambda[w'-w] }{\sum_{u \in \mathcal U} \exp \left[ \xi_u \right] } \right| \\
        &\quad \leq |\mathcal X| (|\mathcal X| - 1) \Lambda_{\mathrm{max}} \lVert w - w' \rVert,
    \end{align*}
    where we used the mean value theorem to find some bounded $\xi \in \mathbb R^{|\mathcal U|}$ between 
    $(r(x,u,\mu_{T-t}) + \Lambda[w])_{u \in \mathcal U}$
    and  
    $(r(x,u,\mu_{T-t}) + \Lambda[w'])_{u \in \mathcal U}$,
    and similarly
    \begin{align*}
        &\lVert F^\alpha(t, w) - G^\alpha(t, w) \rVert \\
        \leq& (L_r \lVert \mu - \mu' \rVert + (|\mathcal X| - 1) L_\Lambda \lVert \mu - \mu' \rVert).
    \end{align*}
    which completes the proof by \cite[Lemma~2]{branicky1994continuity}.

\end{thmproof}

\subsection{System descriptions}
In this section we give more details  on the experiments used for evaluation.
\subsubsection{Left-Right problem}
The LR problem consists of two states $\mathcal{X}=\{L,R\}$, corresponding to left and right, and two actions $\mathcal U =\{S,C\}$ representing staying in the current state or changing position. Choosing action $u=\mathrm{C}$ causes the state to flip between $L$ and $R$ at rate $\beta=0.2$, while choosing $u=\mathrm{S}$ leaves the state unchanged. This leads to the following transition rates:
\[
\Lambda(x,x',u,\mu_t) =
\begin{cases}
\beta, & \text{if } x \neq x', \, u=\mathrm{C}, \\[6pt]
-\beta, & \text{if } x = x', \, u=\mathrm{C}, \\[6pt]
0, & \text{if } u=\mathrm{S}.
\end{cases}
\]
The mean field dependent running reward is defined as
\begin{align*}
    r(x,\mu_t) =
\begin{cases}
-2\,\mu_t(\mathrm{L}), & x=\mathrm{L},\\[4pt]
-\mu_t(\mathrm{R}), & x=\mathrm{R}.
\end{cases}
\end{align*}
This reward function induces an asymmetric swarm-avoidance behavior, penalizing agents more heavily for clustering in the left state than in the right.  The terminal rewards are set to zero, i.e., $q(x)=0$ for all $x\in\mathcal{X}$. In the experiments we set the horizon to $T=50$ and the initial mean field to $\mu_0(\mathrm{L})=0.4$, $\mu_0(\mathrm{R})=0.6$

\subsubsection{Random MFGs}
The random MFGs used for evaluation in Section \ref{sec:exp} consist of $n_x = 10$ states and $n_a = 2$ actions, with a time horizon $T = 10$. The $n_x \times (n_x -1)\times n_a$ mean field independent transition rates and the $n_x \times n_a$ reward table were sampled from a uniform distribution. To promote social distancing, we add the mean field dependent term $\bar r(x,\mu) = -\eta \log (\mu(x))$ to the reward. The terminal rewards are set to zero, i.e., $q(x)=0$ for all $x\in\mathcal{X}$. For the experiments in Section \ref{sec:exp} we choose $\eta = 1.0$. 

\subsubsection{SIS Game}
The SIS game consists of the states $\mathcal{X}=\{S,I\}$ for susceptible and infectious and the actions $\mathcal U =\{N,Q\}$ for no quarantine and quarantine.
The dynamics of the CTMC are defined by the following rates.

\begin{align*}
        \Lambda( x = \mathrm{S},x' = \mathrm{I}, u = \mathrm{N}, \mu_t) &= \kappa \mu(x=\mathrm{I})  \\
        \Lambda(x = \mathrm{S},x' = \mathrm{I}, u_t = \mathrm{Q}, \mu_t) &= 0.0\\
    \Lambda(x = \mathrm{I}, x' = \mathrm{S}  , u = \mathrm{N}, \mu_t) &= \gamma \\
        \Lambda( x = \mathrm{I},x' = \mathrm{S}, u = \mathrm{Q}, \mu_t) &= \gamma,
\end{align*}
where $\gamma = 0.2$ is the healing rate and $\kappa=5.0$ is the infection rate.
The running rewards are given by
\begin{align*}
    r(x=\mathrm{I},u=\mathrm{N},\mu_t) & = - c_i \\
    r(x=\mathrm{I},u=\mathrm{Q},\mu_t) & = - c_i -c_q\\
    r(x=\mathrm{S},u=\mathrm{N},\mu_t) & = 0.0 \\
    r(x=\mathrm{S},u=\mathrm{Q},\mu_t) & = - c_i -c_q,
\end{align*}
where $c_i=10.0$ is the cost for being infected and $c_q=2.0$ is the cost for staying in quarantine. At the final time, infected agents perceive an additional cost of $c_{fi} = 35.0$, yielding 
\begin{align*}
    q(x =\mathrm{I}) &= - c_{fi}\\
    q(x =\mathrm{S}) &= 0.0.
\end{align*}

In the experiments we set the horizon to $T=10$ and the initial mean field to $\mu_0(x=\mathrm{I}) = 0.01 $, $ \mu_0(x=\mathrm{S})=0.99$.



\end{document}